\documentclass[a4paper]{article}

\usepackage{color}
\usepackage{graphicx,amsfonts}
\usepackage{pstricks}

\newcommand{\D}{\mbox{\begin{picture}(0,0)\put(2,0){\framebox(7,7)}\end{picture}}\hspace{.35cm}}

\def\be{\begin{equation}}
\def\ee{\end{equation} }
\def\beq{\begin{eqnarray}}
\def\eeq{\end{eqnarray} }
\def\bc{\begin{center}}
\def\ec{\end{center} }

\def\bit{ \begin{itemize}}
\def\eit{\end{itemize} }

\begin{document}

\title{Estimate for the size of the compactification radius of a one extra dimension Universe}

\author{F. Pascoal$^1$, L.F.A. Oliveira$^2$, F.S.S. Rosa$^3$  and C. Farina$^4$\\ \, \\
\small $^1$ Departamento de F\'isica, Universidade Federal de S\~ao Carlos\\
\small Via Washington Luis, km 235. S\~ao Carlos, 13565-905, SP, Brazil \\
\small $^2$ CENPES, Petr\'oleo Brasileiro S.A.\\
\small  Av. Horácio de Macedo, 950, P 20, S 1082, Ilha do Fundão, RJ,
 21941-915, Brazil.\\
\small $^3$ Theoretical Division, Los Alamos National Laboratory\\
\small Los Alamos, NM, 87544, USA\\
\small $^4$ Instituto de F\'isica, Universidade Federal do Rio de Janeiro\\
\small Caixa Postal 68528, 21941-972, Rio de Janeiro, RJ, Brazil }

\date{\today}

\maketitle

\begin{abstract}
In this work, we use the Casimir effect to probe the existence of
one extra dimension. We begin by evaluating the Casimir pressure
between two plates in a $M^4\times S^1$ manifold, and then use an
appropriate statistical analysis in order to compare the theoretical
expression with a recent experimental data and set bounds for the
compactification radius.
\end{abstract}

\section{Introduction}

In a broad sense, it is fair to say that the search for unification
is the greatest enterprise of theoretical physics. It started a long
time ago, when Sir Isaac Newton showed that celestial and
terrestrial mechanics could be described by the same laws, and
reached one of its highest peaks in the second half of the
nineteenth century, when electricity, magnetism and optics were all
gathered into Maxwell equations.

The quest for unification continued, and, in a historical paper T. Kaluza \cite{Kaluza} managed to combine classical electromagnetism and gravitation into a single, very elegant scheme.
The downside was that his theory required an extra spacial
dimension, for which there was no evidence whatsoever. Some years
later, O. Klein pushed the idea a little further \cite{Klein},
proposing, among other things, a circular topology of a very tiny
radius for the extra dimension, maybe at the Planck scale region.
Although it presented a great unification appeal, the Kaluza-Klein
idea has been left aside for several decades. Only in the
mid-seventies, due to the birth of supergravity theory
\cite{Supergravity}, the extra dimensions came back to the
theoretical physics scenario. As supergravity also had its own
problems, it seemed that the subject would be washed out again, but,
less than a decade later, the advent of string and superstring
theories \cite{Strings} made it a cornerstone in extremely high
energy physics. Nowadays, with the development of M-theory
\cite{M-theory} and some associated ideas, like the cosmology of
branes \cite{branas}, it might even be said that extra dimensions
are almost a commonplace in modern high-energy physics.

\subsection{The hierarchy problem}

If our universe indeed has extra dimensions, then a lot of
intriguing facts should readily come into play. A major development
regarding this issue consists in the alternative approaches to the
so called hierarchy problem, which stands unclear despite all the
efforts carried out over the last thirty years \cite{Peskin}. In
most of the extra-dimensional models, the additional dimensions are
tightly curled up in a small volume, explaining thus how they have
evaded our perception so far. Initially it was thought that {\lq
small volume\rq} should mean {\lq Planck-scale sized volume\rq}, but
now it is conceded that some extra dimensions may be as large as a
human cell, standing at the micrometer scale \cite{refs dim extras}.
Well, how the hierarchy problem fits in this picture? In order to
answer it, let us consider a N-dimensional space-time $R$ in which
$4$ dimensions are large and $n=N-4$ are compactified, containing in
addition a small mass $m$ at a given point $P$. For regions that are
faraway from $P$, at least compared to the compactification radius
$r_c$\footnote{We are tacitly assuming that all the curled up
dimensions are roughly of the same {\lq\lq size\rq\rq},
characterized by $r_c$} everything should be as if the universe were
four-dimensional, so the gravitational interaction is the observed
newtonian field
\begin{equation}\label{potencialgravitacional4}
    {\bf g}_4 = G \frac{m}{r^2} {\bf \hat{r}}= \frac{1}{\left(M_4\right)^2}\frac{m}{r^2} {\bf \hat{r}} \;\; ; \;\;\;\; r \gg
    r_c \;,
\end{equation}
where we took $\hbar=c=1$ and identified the Planck mass $M_{4}
\simeq 1.2 \cdot 10^{19} eV$. When we go to opposite limit ($r \ll
r_c$) it is not possible to ignore the existence of the extra
dimensions anymore, but, from the N-dimensional Gauss law and some
dimensional analysis we get
\begin{equation}\label{leideGauss}
    \int_{S^{n+2}} {\bf g}_N \cdot {\bf dS} = 4 \pi \frac{4 \pi
    m}{\left(M_{N}\right)^{2+n}}\;.
\end{equation}
It is then straightforward to deduce the behavior of the
gravitational field
\begin{equation}\label{potencialgravitacionalN}
{\bf g}_N =
\frac{1}{A_{S^{N-2}}\left(M_{N}\right)^{2+n}}\frac{m}{r^{2+n}} {\bf
\hat{r}} \;\; ; \;\;\;\; r \ll
    r_c \;,
\end{equation}
where $S^{n}$ is the appropriate n-sphere and $A_{S^n}$ stands for
its n-area. Let us notice that we had to choose a tiny n-sphere to
apply the Gauss law, or we would not be in the $r \ll r_c$ regime.
But there is nothing fundamental about this choice, and we may as
well use the Gauss law (\ref{leideGauss}) in order to find the
behavior of the gravitational field for large distances. We shall merely quote
the result
\begin{equation}\label{Pot4PotN}
    {\bf g}_N = \frac{1}{{\left(r_c\right)^n}\left(M_{N}\right)^{2+n}} \frac{m}{r^2} {\bf \hat{r}} \;\; ; \;\;\;\; r \gg
    r_c \;,
\end{equation}
and refer the reader to the bibliography \cite{Dimopoulos} for more
details. Now, comparing (\ref{potencialgravitacional4}) and
(\ref{Pot4PotN}), we find the following constraint relation
\begin{equation}\label{relacaoM4Mn}
   \left(M_4\right)^2 = \left(r_c\right)^n \left(M_{4+n}\right)^{2+n} \;\;\; \Longrightarrow \;\;\; r_c =
   \frac{1}{M_{4+n}} \left(\frac{M_4}{M_{4+n}}\right)^n \;,
\end{equation}
which shows that we may look at $M_4$ as an effective Planck mass,
depending fundamentally on the {\lq true\rq} Planck mass $M_{4+n}$
and the compactification radius $r_c$. That is a very interesting
relation from the perspective of the hierarchy problem, because it
allows for the effective mass $M_4$ that we observe to be huge even
when the true mass $M_{4+n}$ is not that big. Just to put some
numbers, let us consider $r_c \simeq 1 \,\mu m \Rightarrow
\left(r_c\right)^{-1} \!\!\simeq 0.18 \, eV$, which is just below
the lower bound for the experimental validity of newtonian
gravitation \cite{experimentogravitacional}. This
automatically gives the following values for $M_{4+n}$
\begin{eqnarray}\label{listadasmassas}
&&n=1 \;\; \longrightarrow \;\; M_{5} \simeq 0.9 \cdot 10^{7} \; \mbox{TeV} \nonumber \\
&&n=2 \;\; \longrightarrow \;\; M_{6} \simeq 80 \; \mbox{TeV} \\
&&n=3 \;\; \longrightarrow \;\; M_{7} \simeq 110 \; \mbox{GeV} \nonumber
\end{eqnarray}
and so on. Knowing that the electroweak scale $M_{EW}$ is about
$100$ GeV, we conclude that an universe with just one extra
dimension is definitely not the best case scenario. This does not
mean that the reduction of the Planck mass by nine orders of magnitude is
not quite something, but only that a ratio of $M_5$/$M_{EW}$
$\simeq$ $10^8$ still leaves a great {\lq desert\rq} ahead. However,
despite this partial frustration in solving the hierarchy problem,
we will proceed with just one extra dimension, for the plain reason
that it is the simplest model to work with from both the theoretical
and statistical perspectives. Last but not least, it is important to
say that in more sophisticated models it is possible to deal with
the hierarchy issue in a 5-dimensional picture, some of which are
enjoying great success\linebreak nowadays \cite{RandallSundrum}.

\subsection{The Casimir effect}

As the Casimir effect \cite{Casimir48, Reviews} has a strong dependence with
the space-time dimensionality, the Casimir force experiments
\cite{CasimirExperiments, Onofrio} may be a powerful tool to detect the
existence of extra dimensions. In a recent paper, Poppenhaeger et
al. \cite{Poppenhaeger} carried out a calculation in order to set
bounds for the size of an hypothetic extra dimension. They conclude
that their modified expression for the Casimir force with one extra
dimension is with the experimental data of M. Sparnaay
\cite{Sparnaay}, as long as the upper limit for the compactification
radius is at the nanometer range. Although we find these results
very interesting, we would like to stress that the data present in
\cite{Sparnaay} may be inadequate for such estimations, due to its
lack of precision \footnote{As a matter of fact, all that Sparnaay
could conclude from his experiment was that the Casimir force could
not be ruled out, or, in his words: \lq\lq The observed attractions
do not contradict Casimir theoretical prediction.{\rq\rq}
\cite{Sparnaay}.}. It is then a natural step to replace
\cite{Sparnaay} for some more sophisticated experiments, which is
precisely the purpose of this work.

We begin by evaluating the Casimir pressure between two plates in a
hypothetical universe with a $M^4 \times S^1$ topology. We use the
standard mode summation formula for the Casimir effect, and the
calculations are carried out within the analytical regularization
scheme, which is closely related to some generalized zeta functions.
The result for the Casimir energy and pressure show an explicit
dependence on the distance between the plates and on the $S^1$
radius, as they should. As our final task, we use some recent
experimental data \cite{Onofrio} and do a proper statistical
analysis in order to set limits for the values of the
compactification radius.

\section{The Casimir effect in a $M^4 \times S^1$ spacetime}

Let us begin by writing the line element of the $M^4 \times S^1$
universe
\be
ds^2=c^2dt^2-dx^2-dy^2-dz^2-r^2\,d\theta^2,
\ee
where $r$ is the $S^1$ radius. Due to the simplicity of this metric,
the field equations in this manifold are essentially the same as the
minkovskian ones. This holds in particular for the massless
vectorial field, and so we have
\be
\partial_\mu F^{\mu \nu}=0,\,\,\,\,\,\,\,\,\,\,
\partial_\alpha F_{\beta \gamma}+\partial_\beta F_{ \gamma
\alpha}+\partial_\gamma F_{\alpha \beta}=0,
\ee
where
\be
F_{\mu \nu}=\partial_\mu A_{\nu}-\partial_\nu A_{\mu}.
\ee
In the radiation gauge we may write
\be A_0=0,\,\,\,\,\, \partial_\mu A^\mu=0, \label{calibre} \ee
and so the field equation may be recast into
\be
\D A^\mu=0 \label{ecampo}
\ee
Let us assume that the conducting plates are at the planes  $x=0$
and $x=a$. This setup leads to the following boundary conditions
(BC)
\be
F^{\mu\nu} \vert_{x=0} = F^{\mu\nu} \vert_{x=a} = 0 \;\;\;\; \mbox{if}
\;\; \mu\neq1,\,\nu\neq1. \label{cplacas}
\ee
The $S^1$ topology also imposes a periodicity condition for the electromagnetic field
\be
A^{\mu}(x^4)=A^{\mu}(x^4+2\pi r).\label{cperiodica}
\ee
Now we have to solve equation (\ref{ecampo}) constrained by
conditions (\ref{cplacas}) e (\ref{cperiodica}). That is a
straightforward task, so we merely quote the eigenmodes and the eigenfrequencies
\beq
A_1&=&A^{(0)}_1 \cos \left( \frac{m_1\pi x}{a} \right)
\mbox{e}^{i(\vec{k}_{\perp} \cdot \vec{x}_{\perp}+n \theta-\omega
t)},\,\,\,m_1=0,1,2,...\nonumber \\
A_j&=&i\,A^{(0)}_j \sin \left( \frac{m_j\pi x}{a} \right)
\mbox{e}^{i(\vec{k}_{\perp} \cdot \vec{x}_{\perp}+n \theta-\omega
t)},\,\,\,m_j=1,2,...\nonumber \\
\omega^2_{{\bf k}\lambda}&=& \omega^2_{mnk_{\bot}} = \left(\frac{m\pi}{a}\right)^2 +
\left(\frac{n}{r}\right)^2  +  k^2_\perp,\,\,\,\,j=2,3,4;\,\,\,
\,n \in \mathbb{Z}
\eeq
where the fields amplitudes are related by
\be A^{(0)}_1\frac{m_1 \pi}{a}+ {\bf A^{(0)} \cdot k_\perp}
+\frac{n}{r} A^{(0)}_4=0 \, ,
\ee
as a consequence of the gauge condition (\ref{calibre}).

The Casimir energy of the electromagnetic field in a
$M^4 \times S^1$ universe is given by the sum of allowed modes
\beq {\cal E}(a,r) &=&\frac{\hbar}{2} \sum_{{ \bf k} \lambda}
\omega_{{ \bf k} \lambda}  =  \frac{ \hbar c L^2}{8 \pi^2}  \int
d^2{ \bf k}_\parallel \sum_{n=-\infty}^{\infty} \left[ \sqrt{
\left(\frac{n}{r}\right)^2  + k^2_\parallel}  \right.\cr\cr
&+&\left.p \sum_{m=1}^{\infty}
 \sqrt{
\left(\frac{m\pi}{a}\right)^2 + \left(\frac{n}{r}\right)^2 +
k^2_\parallel}\right],\label{energia nao regularizada} \eeq where
$p$ is the number of possible polarizations of the photon ($p=3$ in
this case).
The previous expression is purely formal, since its r.h.s. is
infinite. So, in order to proceed, we introduce a cut-off parameter
$s$ in (\ref{energia nao regularizada}). Then
\beq {\cal E}_{reg}(a,r;s) &=& \frac{L^2 \hbar c}{4 \pi}
\int_0^{\infty} k_\parallel d k_\parallel \sum_{m=-\infty}^{\infty}
\left\{ p \sum_{n=1}^{\infty} \left[ \left(\frac{m\pi}{a}\right)^2 +
\left(\frac{n}{r}\right)^2 +k^2_\parallel
\right]^{\frac{1-s}{2}}\right.\cr\cr &+&\left. \left[
\left(\frac{n}{r}\right)^2 +k^2_\parallel \right]^{\frac{1-s}{2}}
\right\}. \label{energia regularizada} \eeq
Performing the integral in $k_{\parallel}$ we arrive at
\beq {\cal E}_{reg}(a,r;s)&=&  \frac{\hbar c L^2p}{4\pi (s - 3)}
\left(\frac{a}{\pi}\right)^{s-3}\left[ \sum_{m=1}^{\infty} m^{3-s} +
2  \sum_{n,m=1}^{\infty}  \left(  m^2 + \left(\frac{na}{\pi
r}\right)^2 \right)^{\frac{3-s}{2}} \right]\cr\cr &+& \frac{\hbar c
L^2}{2\pi (s - 3)}r^{s-3}\sum_{n=1}^{\infty} n^{3-s}. \label{energia
zeta} \eeq
Let us now recall the definition of the Epstein functions, and, as a particular case, the Riemann zeta
function \cite{Zeta}
\begin{eqnarray}\label{funcoes Epstein}
 E_N(s;a_1,...,a_N)= \sum_{n_1,...,n_N = 1}^{\infty} \left[a_1 n_1^2 + ... + a_N n_N^2 \right]^{-s}, \,\,\,
 \zeta(s)= \sum_{n=1}^{\infty} \frac{1}{n^s} \;.
\end{eqnarray}
By using these definitions, we may recast expression (\ref{energia
zeta}) into
\beq \label{energia zeta 2} {\cal E}_{reg}(a,r;s)&=& \frac{\hbar c
L^2 p}{4\pi (s - 3)}\left(\frac{a}{\pi}\right)^{s-3}
\left[\zeta(s-3) +  2\,
 E_2 \left(\frac{s - 3}{2}; 1, \frac{a^2}{\pi^2 r^2}\right) \right]
 \cr\cr&+& \frac{\hbar c L^2 r^{s-3}}{2\pi(s-3)}\;
\zeta(s - 3)\, . \eeq
The Epstein functions have a well known analytical continuation,
which were thoroughly studied in \cite{Zeta}, among other
references. As a more detailed discussion of that matter would take
us too far afield, let us merely quote the analytic continuation of
the Epstein function $E_2(s;a_1,a_2)$
\beq &&E_2\left(s;a_1^2,a_2^2\right) = - \frac{a_1^{-2s}}{2}
\zeta(2s) + \frac{\sqrt{\pi}}{2 a_2} \frac{\Gamma(s-1/2)}{\Gamma(s)}
a_1^{1-2s} \zeta(2s-1) \cr &+& \frac{2 \pi^{s}}{\Gamma(s)}
\sum_{n_1,n_2=1}^{\infty} a_2^{-s-1/2} \left(\frac{n_1}{a_1
n_2}\right)^{s-1/2} K_{s-1/2} \left( \frac{2 \pi a_1 n_1 n_2}{a_2}
\right) \, , \label{extensao analitica epstein} \eeq
where $K_{\nu}(x)$ stands for the modified Bessel function. The
reflection formula for the Riemann zeta function will also be very
useful
\be
\zeta(s)=\pi^{s-\frac{1}{2}}\frac{\Gamma\left(\frac{1-s}{2}\right)}{\Gamma\left(\frac{s}{2}\right)}\zeta(1-s)\,
. \ee
It is now a straightforward matter put (\ref{energia zeta 2}) into
the form
\beq {\cal E}_{reg}(a,r;s)  &=& p \frac{\hbar cL^2}{4\pi(s - 3)}
\frac{1}{\Gamma\left(\frac{s-3}{2}\right)} \Bigg[\frac{ a^{s -
3}}{\sqrt{\pi}}\Gamma \left(2 - \frac{s}{2}\right)\zeta (4 - s)
\cr\cr &+& \frac{a
r^{s-4}}{\pi^{5-s}}\Gamma\left(\frac{5-s}{2}\right)\zeta(5-s)\cr\cr
&+& \frac{4
a^{\frac{s}{2}-1}}{\sqrt{\pi}}\sum_{m=1}^{\infty}\sum_{n=1}^{\infty}\left(
\frac{mr}{n}\right)^{\frac{s}{2}-2} K_{\frac{s}{2}-2}\left(
\frac{2mna}{r} \right)\Bigg]\cr\cr &+& (p-2)\frac{\hbar cL^2}{4(s -
3)} \frac{\Gamma
\left(2-\frac{s}{2}\right)}{\Gamma\left(\frac{s-3}{2}\right)}\frac{r^{s-3}}{\pi^{\frac{9}{2}-s}}\zeta
(4 - s)\, , \eeq
and, in the limit of $s \rightarrow 0$, we get
\beq {\cal E}(a,r)= &-& \hspace{-7pt} p\frac{\hbar c L^2\pi^2}{1440
a^3} - (p - 2) \frac{\hbar c L^2}{1440 \pi r^3} -  2 p \pi r L^2
\frac{3 \hbar c}{128 \pi^7}\frac{a}{r^5}\zeta(5)\cr\cr &-&
\hspace{-7pt} p\frac{\hbar c L^2}{4\pi^2 r^2
a}\sum_{m=1}^{\infty}\sum_{n=1}^{\infty}\left(
\frac{n}{m}\right)^{2} K_{2}\left( \frac{2mna}{r} \right)
\label{Energia0} \eeq
Due to renormalization issues, we now have to evaluate the Casimir
energy of the region defined by the plates, but with no plates
whatsoever. This calculation is analogous to the one leading to
(\ref{Energia0}), so we merely state the result
\be {\cal E}_{ED}(a,r)= - 2 p \pi r L^2 \frac{3 \hbar c}{128
\pi^7}\frac{a}{r^5}\zeta(5) \, .
 \ee
Then, subtracting this term from (\ref{Energia0}), we finally obtain
the Casimir energy for the $M^4 \times S^1$ with Dirichlet plates
\beq {\cal E}_{Cas}(a,r) = &-& \hspace{-7pt} p\frac{\hbar c
L^2\pi^2}{1440 a^3} - (p-2)\frac{\hbar c L^2}{1440 \pi r^3}\cr\cr&-&
\hspace{-7pt} p\frac{\hbar c L^2}{4\pi^2 r^2
a}\sum_{m=1}^{\infty}\sum_{n=1}^{\infty}\left(
\frac{n}{m}\right)^{2} K_{2}\left( \frac{2mna}{r} \right) \, .
\label{Energia de Casimir} \eeq
The first thing to be stressed about the previous result is that it
precisely coincides with the expression found on
\cite{Poppenhaeger}, although we have derived it in a more clear and
pedagogic way. Now, if we want to make some comparison with the
experiments, we need an expression for the Casimir pressure.
Fortunately, the relation between the Casimir energy and pressure is
a simple one
\beq {\cal P}(a,r) = -\frac{1}{L^2} \frac{\partial {\cal
E}_{Cas}}{\partial a} = &-& \hspace{-7pt} p\frac{\pi^2 \hbar c}{480
a^4} - p\frac{\hbar c}{4\pi^2 r^2
a^2}\sum_{n=1}^{\infty}\sum_{m=1}^{\infty} \left[ 3\left(
\frac{n}{m}\right)^2 K_2\left(\frac{2mna}{r}\right) \right.\cr\cr
&+& \hspace{-7pt} \left.2\frac{n^3 a}{m
r}K_1\left(\frac{2mna}{r}\right) \right] \, , \label{pressao}
 \eeq
where we used some recurrence relations between the modified Bessel
functions \cite{Abramowitz}. If we now make $p=2$ in expressions
(\ref{Energia de Casimir}) and (\ref{pressao}) and take the limiting
case of $r \rightarrow 0$, we will get respectively the standard
Casimir energy and pressure obtained in \cite{Casimir48}.

\section{Estimate of the compactification radius}

The plane geometry is by far the simplest to work with in
theoretical calculations, but unfortunately the situation is not so
friendly from the experimental point of view. A good measurement of the Casimir force between two plates requires, among other things, a high degree of parallelism between the plates, which is very difficult to sustain throughout the course of the experiment. Due to these parallelism problems, the most
popular setup nowadays for measuring the Casimir effect is the sphere-plate configuration \cite{CasimirExperiments}, for which very precise measurements were reported. There is, however, at least one modern experiment designed to detect the Casimir force between parallel plates \cite{Onofrio}, and due to its relevance for us we feel that it is important to describe it a little further.

The apparatus itself used in that experiment is very
interesting. The two parallel plates are simulated by the
opposing faces two silicon beams. One of these beams is rigidly
connected to a frame, in a such a way to provide an accurate control
of the distance between the two beams. The other beam is a thin
cantilever that plays the part of a resonator, since it is free to
oscillate around its clamping point. The apparatus is designed to
measure the square plates oscillating frequency shift ($\Delta
\nu^2$), that is related to the Casimir pressure in the following
way \cite{Onofrio}
\begin{equation}
\Delta \nu^2 =\nu^2 - \nu_0^2 = -\frac{L^2}{4\pi^2 m_{eff}}
\frac{\partial \cal P}{\partial a},
\end{equation}
where $m_{eff}$ is the effective mass of the resonator.

Substituting (\ref{pressao}) in the previous expression, we get
\beq \Delta \nu^2(a,r)&=&-p \frac{\hbar c L^2}{4\pi^2
m_{eff}}\left\{ \frac{\pi^{2}}{120 a^{5}}\right. \cr\cr
&+&\left.\frac{1}{\pi^2 a
r}\sum_{n=1}^{\infty}\sum_{m=1}^{\infty}\left[ \left( 3\,{\frac
{n}{{m}^{3}{a}^{3}}}+\frac{5}{2}\,{\frac
{{n}^{3}}{ma{r}^{2}}}\right)K_{1}\left(\frac{2mna}{r}\right)
\right.\right.\cr\cr &+& \left.\left.\left( 3\,{\frac
{{n}^{2}}{{m}^{2}r{a}^{2}}}+{\frac {{n}^{4}}{{r}^{3}}}\right)
K_{0}\left(\frac{2mna}{r}\right)\right]\right\} \label{ET}. \eeq

Now that we have a theoretical expression of $\Delta \nu^2$ as a
function of $a$ and $r$, we will fit $r$ using the least square
method and the experimental data of \cite{Onofrio}. As we are
fitting just one parameter, we can estimate the best value for $r$
from the graph on figure 1 just by looking for the value of $r$ that
leads to a minimum value of $\chi^2$.

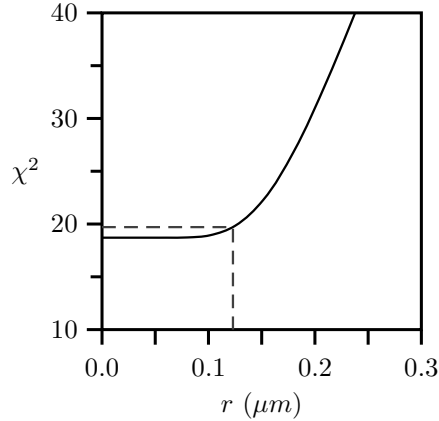
\begin{figure}[!h]
\begin{center}
\newpsobject{showgrid}{psgrid}{subgriddiv=0.5,griddots=10,gridlabels=6pt}
\begin{pspicture}(-1,0.4)(4.2,5.6)
\pscurve [linewidth=.03] (0, 2.61898) (.01750, 2.61898) (.070000,
2.61898) (.1575000000, 2.61898) (.2800000000, 2.61898) (.4375000000,
2.61898) (.6300000000, 2.61898) (.8575000000, 2.61912) (1.120000000,
2.62206) (1.417500000, 2.65034) (1.750000000, 2.77648) (2.117500000,
3.11584) (2.520000000, 3.75788) (2.957500000, 4.69826) (3.430000000,
5.85424)
\pspolygon[fillstyle=solid,fillcolor=white,linestyle=none](0,6)(4.2,6)(4.2,5.6)(0,5.6)
\pspolygon[linewidth=.04](0,1.4)(4.2,1.4)(4.2,5.6)(0,5.6)
\rput(-1,3.5){$\chi^2$}%
\rput(-0.5,1.4){$10$} \psline[linewidth=.04](0,1.4)(-0.2,1.4)%
\psline[linewidth=.04](0,2.1)(-0.15,2.1)%
\rput(-0.5,2.8){$20$} \psline[linewidth=.04](0,2.8)(-0.2,2.8)%
\psline[linewidth=.04](0,3.5)(-0.15,3.5)%
\rput(-0.5,4.2){$30$} \psline[linewidth=.04](0,4.2)(-0.2,4.2)%
\psline[linewidth=.04](0,4.9)(-0.15,4.9)%
\rput(-0.5,5.6){$40$} \psline[linewidth=.04](0,5.6)(-0.2,5.6)%
\rput(2.1,0.4){$r$ ($\mu m$)}%
\psline[linewidth=.04](0,1.4)(0,1.2) \rput(0,0.9){$0.0$}%
\psline[linewidth=.04](0.7,1.4)(0.7,1.2)%
\psline[linewidth=.04](1.4,1.4)(1.4,1.2) \rput(1.4,0.9){$0.1$}%
\psline[linewidth=.04](2.1,1.4)(2.1,1.2)%
\psline[linewidth=.04](2.8,1.4)(2.8,1.2) \rput(2.8,0.9){$0.2$}%
\psline[linewidth=.04](3.5,1.4)(3.5,1.2)%
\psline[linewidth=.04](4.2,1.4)(4.2,1.2) \rput(4.2,0.9){$0.3$}%
\psline[linewidth=.03,linestyle=dashed,linecolor=darkgray](0,
2.75898)(1.722, 2.75898) (1.722,1.4)
\end{pspicture}
\caption{Graph of $\chi^2$ {\it versus} $r$. The vertical dashed line indicates the value $r = 123$ nm, where the function $\chi^2$ hits the value $19.6$ (see footnote 3).} 
\label{fig1}
\end{center}
\end{figure}

Our fit for the compactification radius produced the value of
$0^{+123}_{-0}\ nm$, and the uncertainties on
$r$ give the upper and lower bounds for this radius\footnote{As usual, the
uncertainties were obtained by searching the two values of $r$ that
produce $\chi^2=19.6$ (minimum value plus one). Since a radius can
not be negative, we imposed a vanishing lower bound.}. In a successful fit, the
minimum value of $\chi^2$ should coincide, approximately, with the
number of degrees of freedom of the fit. As in this case we have 8
degrees of freedom\footnote{The degree of freedom of a fit is
defined as being the subtraction of the number of experimental
points used in the fit by the number of adjusted variables. In this
case, we have 9 experimental points and one adjusted variable, which
gives the degree of freedom aforementioned.} and the minimum for
$\chi^2$ turned out to be 18.6, we can state that no good agreement
was obtained between the theoretical model and the experimental
data.

\section{Conclusion}

In this article, we have used the Casimir effect to probe the
existence of one extra dimension. We started by evaluating the
Casimir pressure between two perfect conducting plates living in a
4$+$1 universe, given in (\ref{pressao}), where the extra dimension
is compactified in a $S^1$ topology. In order to set bounds for the
compactification radius, we proceeded to the comparison of this
result with the experimental data of \cite{Onofrio}, and, after an
appropriate statistical analysis, this procedure showed that the
best value for the compactification radius is below approximately 120nm.

We know that the results for the Minkowski space-time are in close
agreement with the experimental data. In
order to be consistent with this picture, the extra compactified
dimension should contribute as a small perturbation to the
four-dimensional result, but, as we have seen, this is not the case.
Among other things, the extra dimension led to a new polarization
degree for the electromagnetic field, which essentially bumped the
$M^4$ result by a factor of approx. $3/2$, that is not small. It is
important to say that this new polarization freedom does not allow
the $r \rightarrow 0$ limit to be taken carelessly, for it
represents the transition from $M^4 \times S^1$ to $M^4$, in which a
polarization degree is discontinuously lost.

We finish by saying that there are other corrections to the Casimir
effect, such as finite conductivity and finite temperature contributions
\cite{Reviews, ThermalEffects}, that we have not taken into account and may completely overwhelm any extra dimensional effects. Besides that, there is the roughness of the plate material
\cite{roughness} and possibly some edge effects \cite{Gies1}, which, if necessary, should also be considered. Hence, in a more rigorous approach, these influences should be taken into
account, and the comparison should be made with very accurate experiments.

\section{Acknowledgments}

F.P. would like to thank CNPq and L.F.A.O. also acknowledges CNPq
for financial support. F.S.S.R. is grateful to both CNPq and FAPERJ
for partial financial support and C.F. thanks CNPq for partial
financial support.


\begin{thebibliography}{99}

\bibitem{Kaluza} T. Kaluza, {\it Zum Unitatsproblem der Physik}, Sitz. Preuss. Akad. Wiss. Phys. Math. {\bf K1} 996 (1921).

\bibitem{Klein} O. Klein, {\it Quantentheorie und funfdimensionale Relativitatstheorie}, Z. Phys. {\bf 37} 895 (1926).

\bibitem{Supergravity} D.Z. Freedman, P. van Nieuwenhuizen and S. Ferrara, Phys. Rev. {\bf D13} 3214
(1976);  E. Cremmer, B. Julia and J. Scherk, Phys. Lett. {\bf B76} 409 (1978).

\bibitem{Strings} M.B. Green, J.H. Schwarz and E. Witten, {\it Supersting Theory}, vols. 1-2,
Cambridge Univ. Press, Cambridge (1987).

\bibitem{M-theory} P. Townsend, hep-th/9612121.

\bibitem{branas} P. Binetruy, C. Deffayet and D. Langlois, Nucl.Phys. B565
269-287 (2000).

\bibitem{Peskin} M. Peskin and D. Schoroeder, {\it An Introduction to Quantum Field
Theory}, Perseus Books Publishing (1995).

\bibitem{refs dim extras} S. Nussinov and R. Schrock, Phys. Rev. {\bf D59}
105002 (1999) and references therein.

\bibitem{Dimopoulos}  N. Arkani-Hamed, S. Dimopoulos and G. Dvali, Phys.Rev. {\bf D59} 086004
(1999).

\bibitem{experimentogravitacional} D.J. Kapner, T.S. Cook, E.G. Adelberger, J.H. Gundlach, B.R. Heckel, C.D. Hoyle and H.E. Swanson, Phys. Rev. Lett. {\bf 98} 021101 (2007).

\bibitem{RandallSundrum} L. Randall and R. Sundrum, Phys. Rev. Lett. {\bf 83} 3370
(1999).

\bibitem{Casimir48}  H.B.G. Casimir, Proc. K. Ned. Akad. Wet. {\bf 51}, 793 (1948).

\bibitem{Reviews} For recent reviews on the Casimir effect, see
M. Bordag, U. Mohideen, and V.M. Mostepanenko, Phys. Rep. {\bf 353}, 1 (2001);
K.A. Milton, J. Phys. A {\bf 24}, R209 (2004);
S.K. Lamoreaux, Rep. Prog. Phys. {\bf 68}, 201 (2005).

\bibitem{CasimirExperiments}
S.K. Lamoreaux, Phys. Rev. Lett. {\bf 78}, 5 (1997); U. Mohideen and A. Roy,
Phys. Rev. Lett. {\bf 81}, 4549 (1998); H.B. Chan, V.A. Aksyuk, R.N. Kleiman, D.J. Bishop and F. Capasso, Science {\bf 291}, 1941 (2001); R.S. Decca, D. Lopez, E. Fischbach, G.L. Klimchitskaya, D.E. Krause and V.M. Mostepanenko, Ann. Phys. {\bf 318} 37 (2005).

\bibitem{Onofrio} G. Bressi, G. Carugno, R. Onofrio and G. Ruoso, Phys. Rev. Lett. {\bf 88}, 041804 (2002); G. Bressi, G. Carugno, A. Galvani, R. Onofrio, G. Ruoso and F. Veronese, Class. Quantum Grav. {\bf 18}, 3943 (2001).

\bibitem{Poppenhaeger} K. Poppenhaeger, S. Hossenfelder, S. Hofmann and M. Bleicher, Phys. Lett. {\bf B582} 1 (2004).

\bibitem{Sparnaay} M.J. Sparnaay, Physica {\bf 24} 751 (1958).

\bibitem{Robin} A. Romeo and A.A. Saharian, J. Phys A: Math. Gen. {\bf 35}, 1297 (2002).

\bibitem{Zeta} E. Elizalde, {\it Ten Physical Applications of Spectral
Zeta Functions}, (Springer-Verlag, Berlin, 1995); see also K.Kirsten, J. Math. Phys. {\bf 35}, 459 (1994).

\bibitem{Abramowitz} M. Abramowitz and I.A. Stegun, {\it Handbook of Mathematical Functions with Formulas, Graphs, and Mathematical Tables}, Dover (1964).

\bibitem{ThermalEffects} I. Brevik, S.A. Ellingsen and K.A. Milton,
New J. Phys. {\bf 8} 236 (2006); V.M.Mostepanenko et al., J.Phys. {\bf A 39} 6589 (2006); G.L. Klimchitskaya and V.M. Mostepanenko, Contemp.Phys. {\bf 47} 131 (2006); M. Brown-Hayes et al., J. Phys. {\bf A} 39, 6195 (2006);  A. Lambrecht, V.V. Nesvizhevsky, R. Onofrio, S. Reynaud, Class.
Quantum Grav. {\bf 22}, 5397 (2005).

\bibitem{roughness} P.A. Maia Neto, A. Lambrecht and S. Reynaud, Phys. Rev. A {\bf 72}, 021115 (2005); C. Genet, A. Lambrecht, P.A. Maia Neto and S. Reynaud, Europhys. Lett. {\bf 62}, 484 (2003); G.L. Klimchitskaya, A. Roy, U. Mohideen and V.M. Mostepanenko, Phys. Rev. A {\bf 60}, 3487 (1999).

\bibitem{Gies1} H. Gies and K. Klingmueller, Phys. Rev. Lett. {\bf 97}
220405 (2006); H. Gies and K. Klingmueller, Phys.Rev. {\bf D74} 045002
(2006).



\end{thebibliography}
\end{document}